\documentclass[structabstract]{aa}  
\usepackage{graphicx}
\usepackage{txfonts}
\usepackage{natbib}
\usepackage{balance}
\usepackage{murl}

\newcommand{\gflog}{\ensuremath{\log gf}}

\newcommand{\teff}{T$_{\rm eff}$}
\newcommand{\glog}{log\,g}
\newcommand{\kms}{\,$\mathrm{km\,s^{-1}}$}

\newcommand{\cobold}{{\sf CO$^5$BOLD}}
\newcommand{\linfor}{{\sf Linfor3D}}

\chardef\ii="10

\idline{13}{13}
\begin{document}
   \title{An upper limit on the sulphur abundance in
 HE\,1327-2326\thanks{based on spectra obtained with CRIRES 
at the 8.2m Antu ESO telescope, programme
386.D-0095} }

%   \subtitle{}

   \author{Piercarlo Bonifacio\inst{1} \and 
Elisabetta Caffau\inst{2,1}\fnmsep\thanks{Gliese Fellow}
\and
Kim A. Venn\inst{3}
\and
David L. Lambert\inst{4}
          }
   \institute{
GEPI, Observatoire de Paris, CNRS, Univ. Paris Diderot; Place
Jules Janssen 92190 Meudon, France \\
              \email{Piercarlo.Bonifacio@obspm.fr}
         \and
    Zentrum f\"ur Astronomie der Universit\"at Heidelberg,
Landessternwarte, K\"onigstuhl 12, 69117 Heidelberg, Germany 
\and
Department of Physics \& Astronomy, 
University of Victoria, Elliott Building, Victoria, BC V8P 5C2, Canada
\and
The W. J. McDonald Observatory, 
The University of Texas at Austin, Austin, 
TX 78712-0259, USA
   \date{Received; accepted }}
% \abstract{}{}{}{}{} 
% 5 {} token are mandatory
  \abstract
  % context heading (optional)
  % {} leave it empty if necessary  
   {Star HE\,1327-2326 is a unique object, with the lowest measured
iron abundance ([Fe/H]$\sim -6$) and a peculiar chemical composition 
that includes large overabundances of C, N, and O 
with respect to iron. One important question
is  whether the chemical abundances in this star reflect the chemical
composition of the gas cloud from which it was formed or if they have
been severely affected by other processes, such as dust-gas winnowing.}
  % aims heading (mandatory)
{We measure or provide an upper limit to the abundance of the volatile
element sulphur, which can help to discriminate between the two scenarios.} 
  % methods heading (mandatory)
   {We observed HE\,1327-2326 with the high resolution infra-red spectrograph
CRIRES at the VLT to observe the \ion{S}{i} lines of Multiplet 3 at 1045\,nm.}
  % results heading (mandatory)
   {We do not detect the \ion{S}{i} line.   A 3$\sigma$ upper limit
on the equivalent width (EW) of any line in our spectrum is EW$<0.66$\,pm.
Using either one-dimensional static or three-dimensional hydrodynamical
model-atmospheres, this translates into a robust upper limit of [S/H]$<-2.6$.}
  % conclusions heading (optional), leave it empty if necessary 
   {This upper limit does not provide conclusive evidence  for or
against dust-gas winnowing, and the evidence coming from other elements 
(e.g., Na and Ti) is also inconclusive or contradictory. 
The formation of dust in the atmosphere versus an origin of the metals
in a metal-poor supernova with extensive ``fall-back'' are not mutually 
exclusive.   It is possible that dust formation distorts the peculiar
abundance pattern created by a supernova with fall-back, thus the 
abundance ratios in HE\,1327-2326 may be used to constrain the properties 
of the supernova(e) that produced its metals, but with some caution.
}

   \maketitle
%
%________________________________________________________________

\section{Introduction}

In the widely accepted cosmological picture, the Universe
experienced a very hot and dense phase (big bang) during which
only the lightest nuclei were synthesised.
All the nuclei, from C to U, were manufactured by stars
in the subsequent evolution of the Universe.
A corollary of this scenario is that the most pristine
stars were formed with a very low content of heavy elements, 
if any at all. 
Low mass stars have very
long lifetimes, in fact stars with masses less than 0.8M\sun\
have Main Sequence lifetimes that exceed the age of the Universe,
as derived from the measurement of the Hubble 
Constant \citep{freedman}.
According to the standard theory of stellar evolution
\cite[see ][for a review on mixing in stars]{pinson}
very minor changes are expected in 
the chemical composition of the atmosphere of a low
mass star during its Main Sequence lifetime. Thus
such stars hold the memory of the chemical composition of the 
interstellar medium from which they formed. 
For this reason, stars with extremely low metallicity are
actively sought  by many research groups 
\citep[an incomplete  list
includes][]{beers85,beers92,cayrel04,christlieb08,cohen,lai,bonifacio09,caffau,bonifacio12} to
address the question of whether low mass stars of primordial (i.e. only H, He, and Li)
chemical composition exist or not \citep[see][for a review on star
formation in primordial stars]{Bromm}.  
In the course of the very successful Hamburg-ESO Survey \citep{christlieb08}
the two stars with the lowest [Fe/H] content so-far measured
were discovered: HE\,0107-5240 \citep{christlieb}, 
a red giant \citep{chris2004},
 and HE\,1327-2326 \citep{frebel}, a subgiant \citep{Korn}.
These two stars show a very remarkable chemical composition 
in which carbon, nitrogen, and oxygen are enhanced by several orders
of magnitude over iron and other iron-peak elements.  
These large overabundances persist even when the analysis
uses 3D hydrodynamical models that imply large
downward revisions of the abundances derived from molecular lines
\citep{collet, Frebel08}.  
In turn, this implies that the global metallicity $Z$
of these stars is not as extreme as implied by their 
iron abundances, e.g., $\log (Z/Z_\odot) \approx -2.5$
for  HE\,1327-2326.
This situation is at odds with what is observed in other
extremely metal-poor stars (see e.g. \citealt{cayrel04,bonifacio09}, and 
\citealt{B_rio} for a review), that show no enhancement of C,
an underabundance of N, and a mild enhancement of O (relative to Fe).
This exceptional situation stimulated a theory
\citep{BL} that actually {\em requires} such an exceptional
chemical composition for low-mass star formation at extremely
low metallicities, i.e., where fine structure lines of \ion{O}{i}
and \ion{C}{ii} provide the cooling necessary to allow
the collapse of a gas cloud of low mass. 
The discovery of SDSS\,J102915+172927 \citep{EC_Nature,EC_AA},
with [Fe/H]$\sim -5$ and no measurable enhancement
of C or N, lies in the ``forbidden zone''
of the \citet{BL} theory, and casts doubt on the
{\em necessity} of a peculiar chemical composition 
at low metallicity. Other means of forming low mass
stars, that rely on fragmentation, either through dust 
\citep{Schneider03,Salvadori,Schneider12,ralf,Schneider12b},
or H$_2$ cooling \citep{clark11,greif11}, appear
viable as alternative routes to star formation and can explain
the existence of stars such as SDSS\,J102915+172927.

The peculiar chemical composition of HE\,0107-5240 and HE\,1327-2326
may arise in several ways, either through enrichment by
at least two supernovae \citep{BonifacioN,Limongi}
or through hypernovae \citep{Umeda,Iwamoto}.

Is the chemical composition of HE\,0107-5240 and  HE\,1327-2326
so exceptional and confined to the extremely metal-poor stars?
In an intriguing paper, \citet{VennLambert}
pointed out that the chemical composition pattern of these two stars,
when plotted as a function of condensation temperature, shows 
similarities to that of post-AGB and $\lambda$\,Boo
stars. Both kinds of objects can be interpreted as the
result of underabundances of elements which
condense onto dust most readily, i.e., at the highest
temperatures, a process dubbed ''dust-gas winnowing''.
In such a scenario,  HE\,0107-5240 and  HE\,1327-2326
could be simply stars affected by dust-gas winnowing
rather than extremely metal-poor stars.
One way to verify such a scenario is to measure the abundance
of volatile elements such as S or Zn. If the extreme
iron deficiency of HE\,0107-5240 and  HE\,1327-2326
is due to dust, we can expect the abundance of the volatile elements
to track more closely that of oxygen, rather than that of iron. 

This paper reports our attempt to measure, or derive
a meaningful upper limit, to the abundance of S
in the brighter of the two stars: HE\,1327-2326.
For Zn, there are already upper limits
from \citet{aoki} ([Zn/Fe]$<  3.07$) using HDS-Subaru spectra
and \citet{Frebel08} ([Zn/Fe] $< 3.01$) using UVES-VLT spectra,
which are too high to examine the claim of dust-gas winnowing, 
and it appears difficult to lower those with existing 
instrumentation.   The available lines of S are stronger 
than those of Zn making the prospect of a detection
more favourable.
Previously existing data covered the \ion{S}{i} lines
of multiplet 1 around 920\,nm, however, that whole region
is severely affected by telluric absorption lines so that
no meaningful upper limit could be derived. 
In this paper, we focus on the lines of multiplet 3 around 
1045\,nm.   Such lines lie in a region that is fairly free 
from telluric absorption and can be usefully used to 
derive sulphur abundances \citep{CaffauS1,CaffauS2}. 

\begin{table}
\caption{Atomic parameters of the sulphur lines, oscillator strengths from
\citet{podobedova}.}
\label{atomic}
\begin{center}
\begin{tabular}{rrrr}
\hline
Wavelength&  Transition               & \gflog  & $\chi _{\rm lo}$    \\
(nm) air  &                                 & &              (eV) \\
\hline
\hline
1045.5449  & $^3\mathrm{S}_1^\mathrm{o}-{^3}\mathrm{P}_2$ &   0.25  & 6.86 \\
1045.6757  & $^3\mathrm{S}_1^\mathrm{o}-{^3}\mathrm{P}_0$ & --0.45  & 6.86 \\
1045.9406  & $^3\mathrm{S}_1^\mathrm{o}-{^3}\mathrm{P}_1$ &   0.03  & 6.86 \\
\hline
\end{tabular}
\\
\end{center}
\end{table}

\section{Observations and results}

CRIRES \citep{crires} at the ESO VLT Antu 8.2\,m telescope 
was used to acquire high resolution spectra of the region 
around the \ion{S}{i} multiplet 3.  The standard setting 
in order 54 centred at 1045.9\,nm with a slit of 0\farcs{4} 
providing a resolving power R$\sim 50\,000$ was used.
Since the star is rather faint (J=12.36) we used long detector  
integration times (DITs) of 300\,s, 6 exposures totalled 4 such DITs,
and one exposure totalled 6 DITs. 
The observations were carried out in service mode using
the standard on-slit nodding (throw of $10''$) to acquire the necessary sky exposures
for sky subtraction.
The spectra themselves, as well as full details on the observations
are available from the ESO archive\footnote{\url{http://archive.eso.org/wdb/wdb/eso/sched_rep_arc/query?progid=386.D-0095(A)}}.
The Adaptive Optic correction was done on-axis on the target itself.
The reduced spectra, using the CRIRES pipeline, were provided by ESO
in the PI-package.
Each spectrum was shifted to wavelength in air, and corrected
by the barycentric correction at the time of observation, 
It was then rebinned at a constant step of
0.004\,nm and the median of the seven spectra was computed.
The signal-to-noise ratio of the combined spectrum is about 67.

   \begin{figure}
   {\centering
   \resizebox{\hsize}{!}{\includegraphics[clip=true]{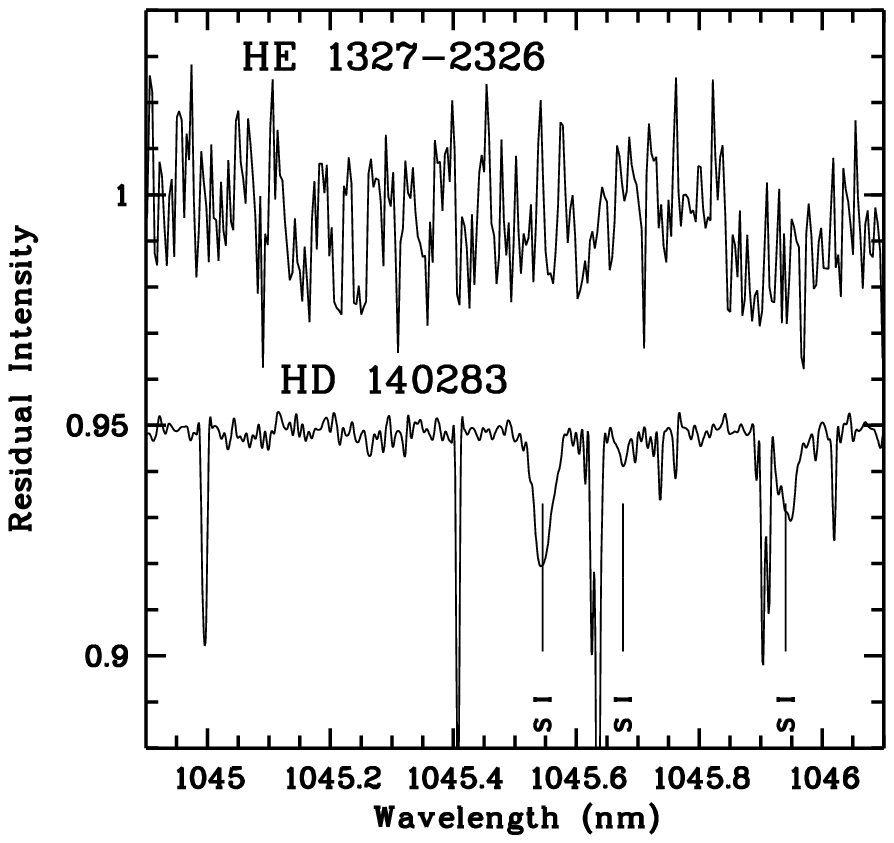}}
}   \caption{The  combined CRIRES spectrum of HE\,1327-2326 
in the spectral region of the \ion{S}{i} Mult.3 lines\label{spec}.
The spectrum has been shifted to zero radial velocity assuming
$v_r=63.8$ \kms \citep{aoki}. Below, shifted by 0.05 units in the vertical
axis, the CRIRES spectrum of HD 140283 \citep{CaffauS2} in which 
all three lines of Mult. 3 are clearly detected, the strongest line has
an equivalent width of about 1.0\,pm, only slightly larger than the
upper limit we determined for HE\,1327-2326. }
\end{figure}

The combined normalised spectrum has been shifted to the
rest wavelength, assuming a radial velocity of 63.8 \kms
\citep{aoki}, and is shown in Fig.\,\ref{spec}. 
No S lines nor telluric lines are apparent.
Using the Cayrel formula \citep{cayrel88}, a 
$3 \sigma$ detection upper limit for this spectrum
is 0.66\,pm.     Attributing this to the strongest line
of the S multiplet, we derive an upper limit to the
the sulphur abundance in HE\,1327-2326 using a model
atmospheres analysis.
The atmospheric parameters for this star are taken 
from \citet{Frebel08}; \teff = 6180\,K, \glog = 3.7 
(c.g.s. units), and microturbulence of 1.6 \kms.
The ATLAS 12 code (\citealt{K05} but see also \citealt{Cast05})
was used to compute a model atmosphere.   For this model, 
the chemical composition found by \citep{Frebel08} was adopted, 
also where the abundances of elements not previously
observed were scaled by $-5.5$ with respect to the solar value, 
or $-5.1$ for the $\alpha$ elements following the general 
trend among metal-poor stars.
For the solar abundance of S we have assumed A(S)=7.16 \citep{abbosun}.
We also have two hydrodynamical simulations, computed
with \cobold\
\citep{Freytag2002AN....323..213F,Freytag2003CO5BOLD-Manual,Wedemeyer2004A&A...414.1121W,freytag12}, 
and their associated 
1D LHD models \citep{cl07}, although
their chemical composition is standard.
We used the atomic data provided in Table \ref{atomic}
and the \linfor \footnote{http://www.aip.de/$\sim$mst/Linfor3D/linfor\_3D\_manual.pdf} 
code to compute the S abundance implied
by the upper limit on the equivalent width.
The results are provided in Table \ref{ulims}, 
which also includes the NLTE correction deduced from the
tables of \citet{Takeda}, assuming metallicity --3.0
(this metallicity is preferred to avoid extrapolation).  
Our observation provides a robust upper limit of [S/H] $< -2.60$,
independent of the choice of model atmosphere and line formation code.

   \begin{figure}
   {\centering
   \resizebox{\hsize}{!}{\includegraphics[clip=true]{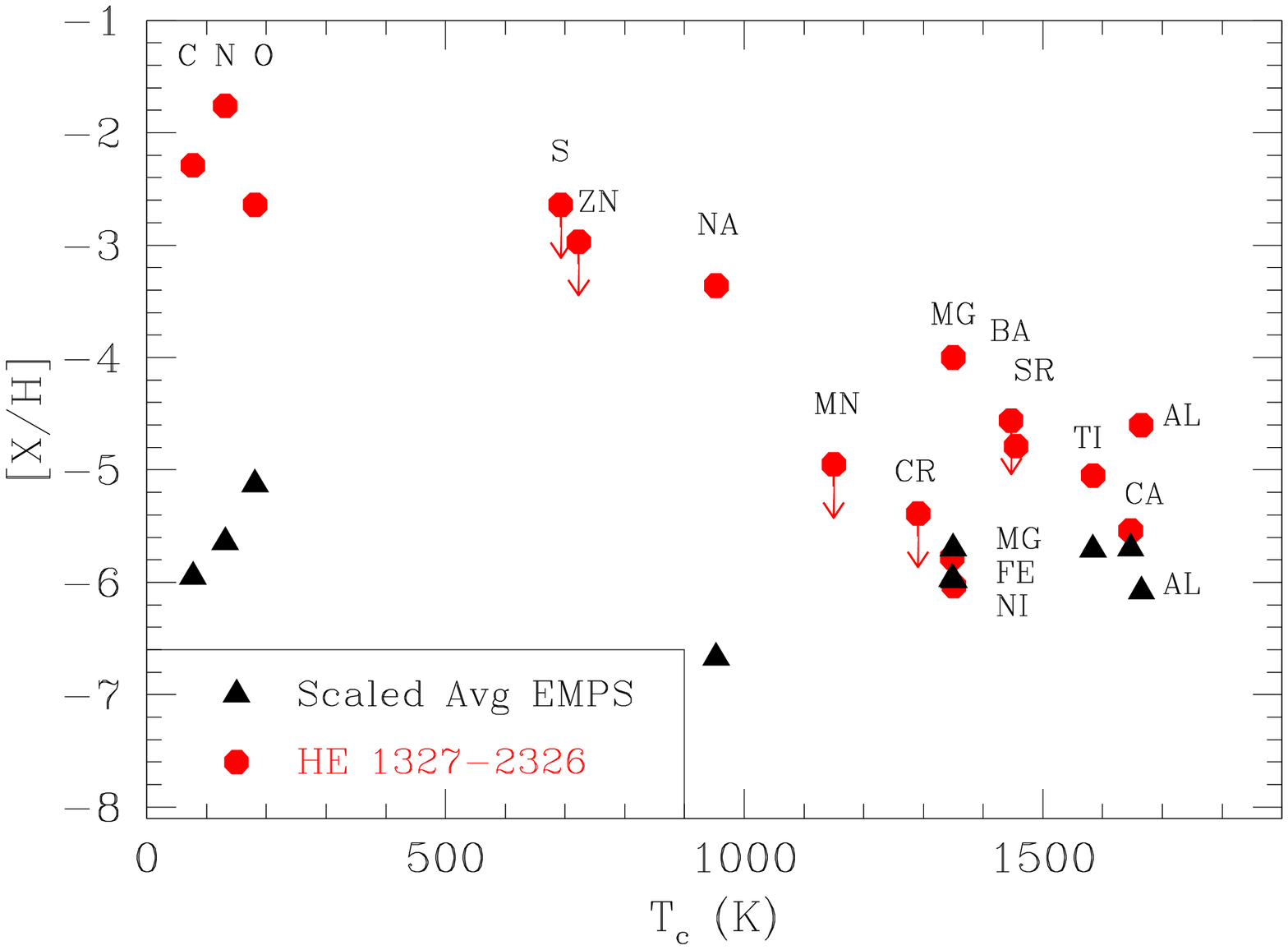}}
}   \caption{\label{plavgs}  Abundances in  HE\,1327-2326 
(red circles) as a function of condensation temperature
(Lodders 2003)
compared to those of the ``average star'' (black triangles),
as defined by \citet{LC12} starting from the
data of \citet{cayrel04} and \citet{spite05}.
The symbols for Fe and Ni in the ``average star'' are on top
of one another.
}
\end{figure}

\begin{table}
\caption{\label{ulims}
Upper limits on  the S abundance in  HE\,1327-2326}
\begin{center}
\begin{tabular}{lrrrrr}
\hline
Model & \teff & \glog & [M/H] & A(S)& NLTE corr \\
\hline
\hline
ATLAS 12 & 6180 & 3.7 & $-5.5$ & $<4.72$ & -0.3\\
\cobold  & 6300 & 4.0 & $-3.0$ & $<4.87$ & -0.3\\
\cobold  & 6300 & 3.5 & $-3.0$ & $<4.68$ & -0.3 \\ 
\hline
\end{tabular}
\end{center}
\end{table}

   \begin{figure}
   {\centering
   \resizebox{\hsize}{!}{\includegraphics[clip=true]{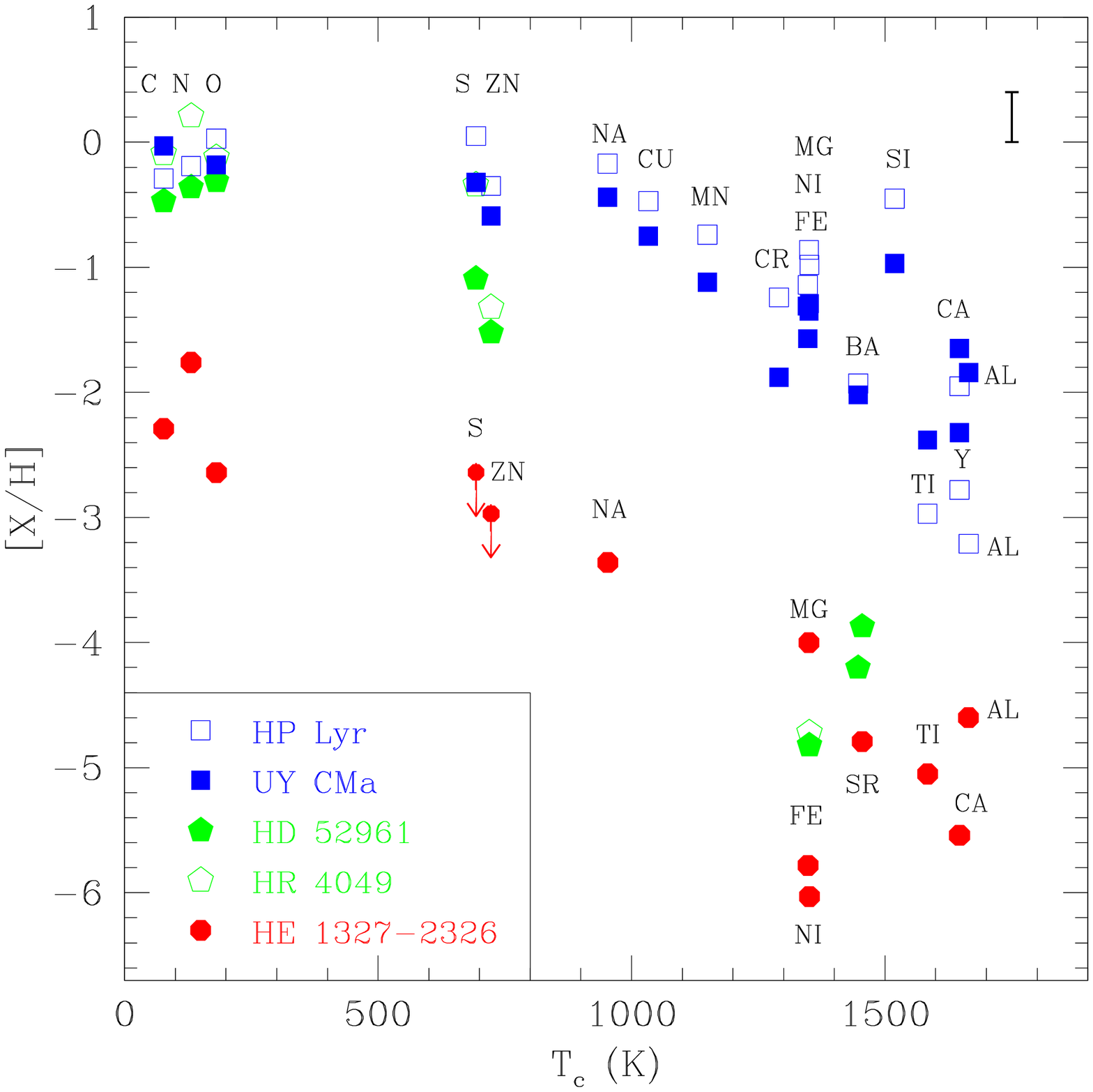}}
}   \caption{\label{fig3}  Elemental abundances for
HE\,1327-2326, two post-AGB stars, and RV Tauri stars 
plotted versus condensation temperature $T_C$ from 
\citet{Lodders} for the solar composition.
HP Lyr and UY CMa are RV Tauri stars, the abundances
are from \citet{giridhar}. HD 52961 and HR 4049 are A-type post-AGB 
stars and the abundances are from \citet{waelkens}, plus
Zn  from \citet{takeda02}, and 
\citet{vanwinckel}, respectively. 
This is a slightly modified version of Fig. 2 of \citet{VennLambert}.
 }
\end{figure}

\section{Discussion}

Our upper limit to the S abundance, [S/H] $< -2.6$ or [S/Fe] $< 3.4$, 
is the tightest upper limit obtainable for HE1327-2326 due to our selection
of \ion{S}{i} lines.   This limit has two possible interpretations given 
the overall composition of this extremely iron-poor star \citep{Frebel08}:
(i) the star was formed from primordial gas contaminated by first-generation
supernovae only, or (ii) the composition is severely affected
by dust-gas winnowing,
i.e., elements which condense into and onto dust grains at moderate temperatures 
are underabundant with respect to their initial stellar abundances. 

\citet[][Figure 17]{Frebel08} summarise the predicted abundances from several
models for first-generation supernovae. 
Several of these models provide 
a reasonable account of the observed abundances of C, N, O, Na, Mg, Al, and  
Ca to Ni. In the case of S,  the three models providing a
prediction give [S/Fe] from 1.9 to 0.9  which are, of course, consistent
with our observed limit  [S/Fe] $< 3.4$. Given the nature of nucleosynthesis
by supernovae from massive stars, the [S/Fe] seems certain to be bracketed by
the observed [Mg/Fe] and [Ca/Fe], i.e., 2 to 1 in this star. 
Other measurements of S in stars 
are not particularly helpful because they do 
not extend to ultra-metal poor stars;
for example, [S/Fe] $\simeq +0.3$ 
for metal-poor stars down to [Fe/H] $\simeq -3.5$
\citep{nissen07,spite11}.
In summary, the limit [S/Fe] $<3.4$ is too loose a value to contradict the
suggestion  that HE 1327-2326 formed from gas contaminated by first-generation
supernovae. 

According to the radical alternative viewpoint, 
HE 1327-2326's composition may be
seriously affected by dust-gas winnowing. A signature of winnowing is that
the underabundance of an element is correlated with its predicted
condensation temperature. Elements with the lowest condensation
temperatures are the surest indicators of a star's initial metallicity.
In this regard, C, N, and O suggest a 
metallicity much greater than the [Fe/H] = $-6$  
obtained from a set of Fe\,{\sc i} lines. 

   \begin{figure}
   {\centering
   \resizebox{\hsize}{!}{\includegraphics[clip=true]{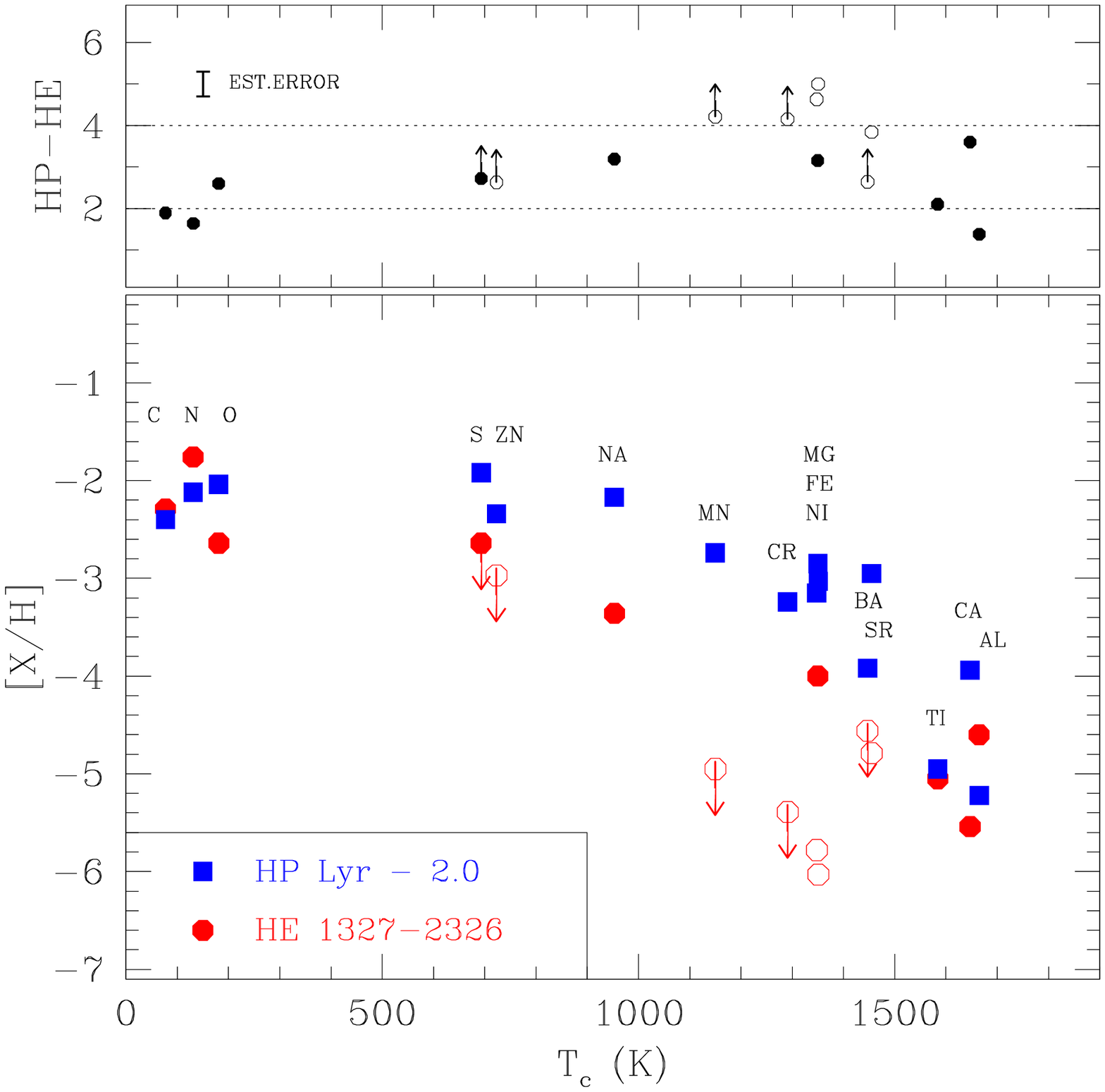}}
}   \caption{\label{fig4}  Abundances in HE\,1327-2326 compared
to those of the RV Tauri star HP Lyr, after the latter have 
been scaled by --2\,dex. 
For HE\,1327-2326 filled symbols are used for elements with 
$A\le 22$ and open symbols the heavier elements.
The upper box shows the differences in the abundances
scaled HP\,Lyr -- HE\,1327-2326. 
 }
\end{figure}

In Fig.\,\ref{plavgs}, we plot the abundances of   
HE\,1327-2326 as a function of condensation temperature $T_C$
predicted by \citet{Lodders} for the solar
composition at low gas pressure,
and compare it to that of the ``average'' metal-poor star.
This is 
defined by \citet{LC12}, starting from the data
of \citet{cayrel04} and \citet{spite05}.
The sample of \citet{cayrel04} spans the metallicity
range --2.5 to --4.0.
The fact that the pattern of this 
``average star'' closely follows that of
 SDSS\,J102915+172927 \citep{EC_Nature,EC_AA} implies 
that this pattern extends down to [Fe/H]$=-5.0$.
To make the comparison in Fig.\,\ref{plavgs}, we assume 
that stars with this pattern exist down to [Fe/H]=--6.0.
The figure highlights the fact that the abundance pattern 
in the ``average'' metal poor star shows no trend with 
$T_C$, while HE\,1327-2326 shows a 
very definite correlation.  The upper limits for S and Zn 
fall close to the mean trend for elements with $T_C < 1000$ K.

The chosen condensation temperatures were computed for a solar 
C/O ratio (i.e., 0.5) but the observed C/O is non-solar (i.e., 2.3) 
--  a change which will require adjustments to the $T_C$ values.
It would be helpful to calculate $T_C$ 
for a range of compositions more representative of the possible
initial compositions of HE\,1327-2326. 
This may result in a smoother trend of abundances versus $T_C$.  
In a different context, an attempt to derive chemical compositions
of planets hosted by stars of different chemical composition, 
\citet{Bond} have computed the $T_C$ per element for the chemical 
compositions of different planet host stars.
The differences in the C/O ratios can be quite significant, 
e.g., 300\,K.  Although this result is unlikely to apply 
directly to the low metallicity star HE\,1327-2326, it suggests
that significant variations due to chemical composition are 
possible.

In Fig. \ref{fig3}  we plot [X/H] of HE\,1327-2326
versus the condensation temperature $T_C$, and
in the same plot are shown the abundances
of several comparison stars that are known to show the
effects  of dust-gas winnowing: post-AGB and RV Tauri stars.
These post-AGB and RV Tauri stars show IR excesses as 
evidence of warm dust \citep{lamers1986, waelkens, gielen2008, 
friedemann1996, deruyter2005}.
In spite of the general resemblance in the chemical pattern,
HE\,1327-2326 shows considerable scatter among the C, N, and O 
abundances at low $T_C$ and again at high $T_C$, 
e.g., Mg relative to Fe and Ni.

In Fig. \ref{fig4} we  plot the abundances as a function of $T_C$
for HE\,1327-2326 together with those of the RV Tauri star HP\,Lyr 
scaled by --2 dex.  The mismatch of Mg with Fe and Ni is obvious. 
What is also striking is that only a subset of the heavier elements 
($A > 22$ shown as open symbols) show large deviations from the 
scaled pattern of HP\,Lyr, i.e., only the iron-group elements
(Fe and Ni, and the upper limits on Cr and Mn) differ from the
dust-gas winnowing pattern.   The other heavy elements (Sr,
and the upper limits on Ba and Zn) are consistent with pattern
in HP Lyr.

   \begin{figure}
   {\centering
   \resizebox{\hsize}{!}{\includegraphics[clip=true]{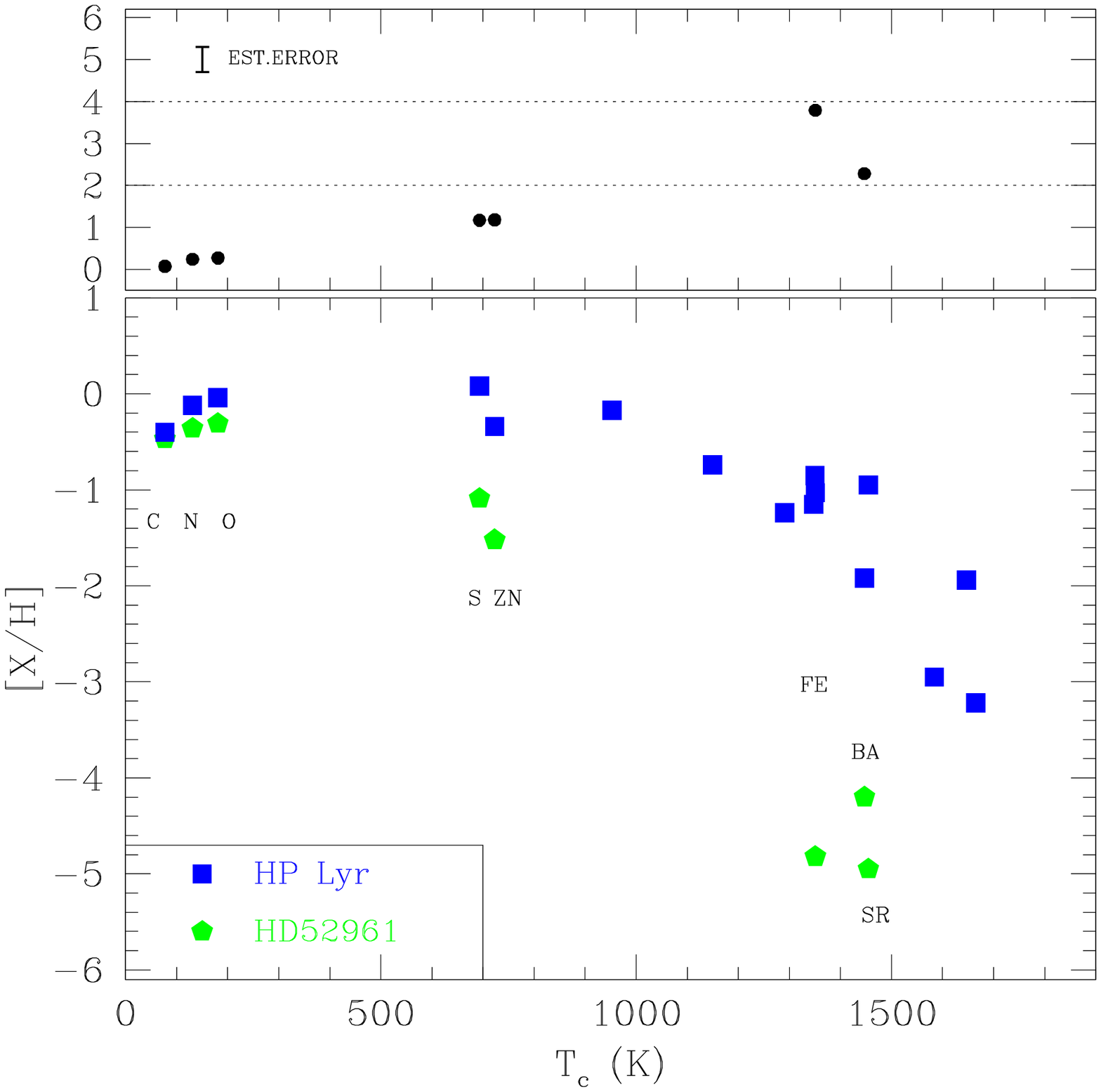}}
}   \caption{\label{fig5} Abundances in the post-AGB star HD\,52961
compared to those in the RV Tauri star HP\,Lyr.
The upper box shows the differences in the abundances
HP\,Lyr -- HD\,52961. 
 }
\end{figure}

A similar exercise is shown in the Fig.\,\ref{fig5} where
the abundances of HP\,Lyr and the post-AGB star HD\,52961 
are compared.  
This figure suggests, as do studies 
of RV Tauri variables, that dust-gas winnowing  does not
result in a universal abundance pattern with $T_C$.
The properties of a debris disk would depend on
composition, age, distance from the host star, 
and the dust temperature.
One should however also be aware that we are 
comparing the patterns of stars that occupy
different positions in the Hertzsprung-Russel diagram.
In particular  the mass of the outer convection zone
should be revelant to the final abundance pattern
observable in the photosphere in the
presence of dust-gas winnowing.

A complication is that we do not know what the
``intrinsic'' abundances in HE\,1327-2326 would be 
{\it before distortion by dust-gas winnowing}
if this star has been affected by the presence of dust. 
 While it is tempting to assign the [O/H] value as the
intrinsic metallicity (i.e., [X/H] = --2), this is
complicated by the type(s) of SNe that would have
enriched the proto stellar nebula.    
If the abundance pattern of
HE\,1327-2326 is a signature of nucleosynthesis
in the supernovae of the earlier generation, it seems
unavoidable to invoke ``fall-back'' in at least one
of  the  supernovae, in order to account for the very high
O/Fe ratio. This is the case also for  
HE\,0107-5240  \citep{BonifacioN,Limongi,Umeda,Iwamoto}.
The ``fall-back'' increases when the core is more compact owing 
to a stronger reverse shock.  This tends to occur in the Pop III 
models that result in very compact black holes, whereas the
Pop I (solar metallicity) models do not \citep{zhang},
 although there are other factors to consider 
such as explosion energy, rotation rate, and progenitor mass
\citep{Joggerst}.  
Thus, the specific role of metallicity and thus Pop I/III progenitor 
is unclear, e.g., \citet{moriya10} have shown that their solar metallicity
core-collapse supernovae models can result in significant ``fall-back'',
used to explain the light curve and lack of metal absorption lines 
in the spectrum of the peculiar supernova SN 2008ha.

Referring to  Fig. 17 of \citet{Frebel08},  
the general trend with atomic number of the
model prediction is by and large similar to the observed
one. Unfortunately the pattern produced
by ``fall-back'' is also similar
to what would be the effect of dust formation.
The exceptions are  Na, S and Ti. 
Sodium is only moderately depleted onto dust
and is in fact found at the same level as CNO and S
for the scaled abundances of HP\,Lyr in Fig.\,\ref{fig4}.
In metal-poor stars Na is found to be slightly 
underabundant with respect to iron,  0.2 -- 0.3\,dex
\citep{sergei}. 
On the other hand all the SN models underproduce Na with 
respect to N by at least 1\,dex. 
The fact that 
in Fig.\,\ref{fig4} the Na abundance in  HE\,1327-2326
is about 1\,dex below the scaled abundance of HP\,Lyr
suggests that the result is due to nucleosynthesis, 
not of dust formation. 
Sulphur is predicted to be
roughly at the same level as Ca in the SNe products, but 
our upper limit is inconclusive, being consistent with
[S/Ca]$\sim 0$ as expected from the SN nucleosynthesis
and [S/Ca] $< 0$ as expected from dust formation. 
Always with reference to Fig.\,17 of \citet{Frebel08},
titanium is expected to be underproduced in SNe with
respect to Ca, by about 1\,dex. The observed
abundance of Ti in  HE\,1327-2326 is higher than 
all the model predictions.  In Fig.\,\ref{fig4}
we may see the the Ti abundance almost matches that
of HP\,Lyr scaled by --2.0\,dex. This would 
therefore suggest that we are observing the effects of dust
and not of nucleosynthesis.

There  remains one significant difference
between the comparison in Fig.\,\ref{fig4} and that in Fig.\ref{fig5}.
In Fig\,\ref{fig5} the abundances [X/H] of the CNO elements, 
presumably unaffected by dust,
are very similar. Instead in Fig.\,\ref{fig4}, the pattern is  different.
Possibly the most remarkable fact is that in HE\,1327-2326 oxygen
displays the lowest abundances, while it is the largest in HP\,Lyr.
The pattern in CNO abundances found in extremely
metal poor stars at [Fe/H]$\sim -3.5$ is       
[O/H]$\sim -2.5$ \citep[][unevolved stars]{jonay},
[C/H]$\sim -3.5$ \citep[][for ``unmixed'' giants]{spite05} and 
[N/H]$\sim -3.5$ \citep[][for ``unmixed'' giants]{spite05},
which is clearly different from that observed in HE\,1327-2326.
Therefore even if the abundances of other elements were affected
by dust-gas winnowing, the abundances of C, N, and O are 
peculiar, with respect to those of other extremely metal-poor stars.   

In an approach based on theoretical models of Galactic chemical
evolution, we may consider adjusting to the ``intrinsic'' N and O 
abundances of \citet{GP}, however this widens the gap between  
the N and O abundances  to about 1.2 dex. 
At high $T_C$ in Fig.\,\ref{fig3}, corrections for the initial
abundances of the $\alpha$-elements Mg, Ca, and Ti would marginally 
reduce the spread because [$\alpha$/Fe] $\simeq +0.4$ initially.
However, Al remains a clear outlier because [Al/Fe] $\simeq 0$ initially 
\citep{GP,chiaki}.

Amongst the mechanisms that may have distorted the abundances of 
HE\,1327-2326, one should also consider atomic diffusion. 
This has been considered in detail by \citet{Korn}, who concluded,
on the basis of parametric models of diffusion in the
presence of  turbulence, that the effects are small, of the
order of 0.2\,dex. 
Pure diffusion, not inhibited by turbulence, would
produce effects much larger than this. 
It is not inconceivable that 
some peculiar phenomenon takes place 
in connection with atomic diffusion at the 
lowest metallicities, e.g., one peculiar
behaviour found at the lowest metallicities is
the meltdown of the Spite plateau below [Fe/H]=--3
\citep{sbordone}.

\balance

\section{Conclusions}

Our upper limit to the S abundance does not provide decisive 
evidence in favour of the formation of dust in the atmosphere
of HE\,1327-2326, but this cannot be ruled out either. 
The pattern in the light element abundances as a function
of condensation temperature still suggests that 
HE\,1327-2326's atmosphere could have been distorted by 
dust-gas winnowing.   The abundances resulting from a SN 
experiencing extensive  ``fall-back'' would show a similar 
pattern though.   Thus, interpretation of the abundance 
pattern is not unambiguous. 
Two measured elements that could help to discriminate
between the two scenarios, Na and Ti, provide
contradictory results. 
Na/N supports a nucleosynthetic origin for the 
abundance pattern, on the contrary Ti 
is in better agreement with the dust hypothesis.
Nucleosynthesis and dust-gas winnowing are not mutually 
exclusive (e.g., \citealt{Schneider12}), 
and it is possible that dust formation distorts the peculiar 
abundance pattern created by a supernova with extensive
``fall-back''.  The available information, albeit contradictory,  
suggests that any diagnostics for supernovae yields based on 
abundance ratios in this star be treated with caution.

\begin{acknowledgements}
We are grateful to F. Herwig, A. Heger and M. Pignatari for useful
discussions on nucleosynthesis in low metallicity supernovae.
We wish to thank the referee, A. Korn, for his thoughtful comments
that helped to improve the paper.
We acknowledge support from the Programme National
de Physique Stellaire (PNPS) and the Programme National
de Cosmologie et Galaxies (PNCG) of the Institut National de Sciences
de l'Univers of CNRS.   KAV also thanks NSERC for support through the
Discovery Grant program.
\end{acknowledgements}


\begin{thebibliography}{}

\bibitem[Andrievsky et 
al.(2007)]{sergei} 
Andrievsky, S.~M., Spite, M., Korotin, S.~A., et al.\ 2007, \aap, 464, 1081 




\bibitem[Aoki et al.(2006)]{aoki} Aoki, W., Frebel, A., 
Christlieb, N., et al.\ 2006, \apj, 639, 897 


\bibitem[Beers et al.(1985)]{beers85} Beers, T.~C., Preston, 
G.~W., \& Shectman, S.~A.\ 1985, \aj, 90, 2089

\bibitem[Beers et al.(1992)]{beers92} Beers, T.~C., Preston, 
G.~W., \& Shectman, S.~A.\ 1992, \aj, 103, 1987 

\bibitem[Bond et al.(2010)]{Bond} Bond, J.~C., O'Brien, 
D.~P., \& Lauretta, D.~S.\ 2010, \apj, 715, 1050 


\bibitem[Bonifacio(2010)]{B_rio} Bonifacio, P.\ 2010, IAU 
Symposium, 265, 81 


\bibitem[Bonifacio et al.(2003)]{BonifacioN} Bonifacio, P., 
Limongi, M., \& Chieffi, A.\ 2003, \nat, 422, 834 

\bibitem[Bonifacio et 
al.(2009)]{bonifacio09} Bonifacio, P., et al.\ 2009, \aap, 501, 519 

\bibitem[Bonifacio et al.(2012)]{bonifacio12} Bonifacio, P., 
Sbordone, L., Caffau, E., et al.\ 2012, \aap, in press, arXiv:1204.1641 




\bibitem[Bromm 
\& Loeb(2003)]{BL} Bromm, V., \& Loeb, A.\ 2003, \nat, 425, 812 

\bibitem[Bromm 
\& Larson(2004)]{Bromm} Bromm, V., \& Larson, R.~B.\ 2004, \araa, 42, 79 






\bibitem[Caffau 
\& Ludwig(2007)]{cl07} Caffau, E., \& Ludwig, H.-G.\ 2007, \aap, 467, L11 

\bibitem[Caffau et 
al.(2007)]{CaffauS1} Caffau, E., Faraggiana, R., Bonifacio, P., Ludwig, H.-G., \& Steffen, M.\ 2007, \aap, 470, 699 

\bibitem[Caffau et al.(2010)]{CaffauS2} Caffau, E., Sbordone, 
L., Ludwig, H.-G., Bonifacio, P., 
\& Spite, M.\ 2010, Astronomische Nachrichten, 331, 725 




\bibitem[Caffau et al.(2011a)]{abbosun} Caffau, E., Ludwig, 
H.-G., Steffen, M., Freytag, B., \& Bonifacio, P.\ 2011a, \solphys, 268, 255

\bibitem[Caffau et 
al.(2011b)]{caffau} Caffau, E., Bonifacio, P., Fran{\c c}ois, P., et al.\ 2011b, \aap, 534, A4 

\bibitem[Caffau et al.(2011c)]{EC_Nature} Caffau, E., Bonifacio, 
P., Fran{\c c}ois, P., et al.\ 2011c, \nat, 477, 67 

\bibitem[Caffau et 
al.(2012)]{EC_AA} Caffau, E., 
Bonifacio, P., Fran{\c c}ois, P., et al.\ 2012, \aap, 542, A51 


\bibitem[Castelli(2005)]{Cast05} Castelli, F.\ 2005, Memorie 
della Societ\`a Astronomica Italiana Supplementi, 8, 25 



\bibitem[Cayrel(1988)]{cayrel88} Cayrel, R.\ 1988, in 
IAU Symp. 132 The Impact of  Very High S/N
Spectroscopy on Stellar Physics, G. Cayrel de Strobel and M. Spite eds.,
p. 345 



\bibitem[Cayrel et 
al.(2004)]{cayrel04} Cayrel, R., et al.\ 2004, \aap, 416, 1117


\bibitem[Cohen et al.(2008)]{cohen} Cohen, J.~G., Christlieb, 
N., McWilliam, A., et al.\ 2008, \apj, 672, 320 


\bibitem[Christlieb et al.(2002)]{christlieb} Christlieb, N., et 
al.\ 2002, \nat, 419, 904 

\bibitem[Christlieb et al.(2004)]{chris2004} Christlieb, N., 
Gustafsson, B., Korn, A.~J., et al.\ 2004, \apj, 603, 708 



\bibitem[Christlieb et 
al.(2008)]{christlieb08} Christlieb, N., Sch{\"o}rck, T., Frebel, A., 
Beers, T.~C., Wisotzki, L., \& Reimers, D.\ 2008, \aap, 484, 721

\bibitem[Clark et al.(2011)]{clark11} Clark, P.~C., Glover, 
S.~C.~O., Smith, R.~J., et al.\ 2011, Science, 331, 1040 

\bibitem[Collet et al.(2006)]{collet} Collet, R., Asplund, M., 
\& Trampedach, R.\ 2006, \apjl, 644, L121 



\bibitem[de Ruyter et al.(2005)]{deruyter2005} de Ruyter S., 
van Winckel H., Dominik C., Waters L.B.F.M., Dejonghe H., 2005,
\aap, 435, 161






\bibitem[Frebel et al.(2005)]{frebel} Frebel, A., et al.\ 
2005, \nat, 434, 871 



\bibitem[Frebel et al.(2008)]{Frebel08} Frebel, A., Collet, R., 
Eriksson, K., Christlieb, N., \& Aoki, W.\ 2008, \apj, 684, 588 

\bibitem[Freedman et al.(2001)]{freedman} Freedman, W.~L., 
Madore, B.~F., Gibson, B.~K., et al.\ 2001, \apj, 553, 47 

\bibitem[Friedemann et al.(1996)]{friedemann1996} Friedemann, C., 
Guertler J., Loewe M., 1996, \aaps, 117, 205


\bibitem[{{Freytag} {et~al.}(2002){Freytag}, {Steffen}, \&
  {Dorch}}]{Freytag2002AN....323..213F}
{Freytag}, B., {Steffen}, M., \& {Dorch}, B. 2002, Astronomische Nachrichten,
  323, 213

\bibitem[{{Freytag} {et~al.}(2003){Freytag}, {Steffen}, {Wedemeyer-B{\"o}hm},
  \& {Ludwig}}]{Freytag2003CO5BOLD-Manual}
{Freytag}, B., {Steffen}, M., {Wedemeyer-B{\"o}hm}, S., \& {Ludwig}, H.-G.
  2010, {CO5BOLD User Manual},
  \verb|http://www.astro.uu.se/~bf/co5bold_main.html|


\bibitem[Freytag et al.(2012)]{freytag12} Freytag, B., Steffen, 
M., Ludwig, H.-G., et al.\ 2012, Journal of Computational Physics, 231, 919 


\bibitem[Gielen et al.(2008)]{gielen2008} Gielen C., van Winckel H, Min M.,
 Waters L.B.F.M. \& Lloyd Evans T. \ 2008, \aap, 490, 725 

\bibitem[Giridhar et al.(2005)]{giridhar} Giridhar, S., Lambert, 
D.~L., Reddy, B.~E., Gonzalez, G., \& Yong, D.\ 2005, \apj, 627, 432 

\bibitem[Gonz{\'a}lez Hern{\'a}ndez et 
al.(2010)]{jonay} Gonz{\'a}lez Hern{\'a}ndez, J.~I., 
Bonifacio, P., Ludwig, H.-G., et al.\ 2010, \aap, 519, A46 


\bibitem[Goswami 
\& Prantzos(2000)]{GP} Goswami, A., \& Prantzos, N.\ 2000, \aap, 359, 191 


\bibitem[Greif et al.(2011)]{greif11} Greif, T.~H., Springel, 
V., White, S.~D.~M., et al.\ 2011, \apj, 737, 75 

\bibitem[Iwamoto et al.(2005)]{Iwamoto} Iwamoto, N., Umeda, H., 
Tominaga, N., Nomoto, K., \& Maeda, K.\ 2005, Science, 309, 451 

\bibitem[Joggerst et al. (2010)]{Joggerst} Joggerst C.C., Almgren A.,
Bell J., Heger A., Whalen D., Woosley S.E., 2010, \apj, 709, 11


\bibitem[Kaeufl et al.(2004)]{crires} Kaeufl, H.-U., 
Ballester, P., Biereichel, P., et al.\ 2004, \procspie, 5492, 1218 

\bibitem[Klessen et al.(2012)]{ralf} Klessen, R.~S., Glover, 
S.~C.~O., \& Clark, P.~C.\ 2012, \mnras, 421, 3217 

\bibitem[Kobayashi 
\& Nakasato(2011)]{chiaki} Kobayashi, C., \& Nakasato, N.\ 2011, \apj, 729, 16 

\bibitem[Korn et al.(2009)]{Korn} Korn, A.~J., Richard, O., 
Mashonkina, L., et al.\ 2009, \apj, 698, 410 


\bibitem[Kurucz(2005)]{K05} Kurucz, R.~L.\ 2005, Memorie 
della Societ\`a Astronomica Italiana Supplementi, 8, 14 

\bibitem[Lamers(1986)]{lamers1986} Lamers H.J.G.L.M, 1986 \aap, 159, 90

\bibitem[Lai et al.(2008)]{lai} Lai, D.~K., Bolte, M., 
Johnson, J.~A., et al.\ 2008, \apj, 681, 1524 

\bibitem[Limongi et al.(2003)]{Limongi} Limongi, M., Chieffi, 
A., \& Bonifacio, P.\ 2003, \apjl, 594, L123 

\bibitem[Limongi 
\& Chieffi(2012)]{LC12} Limongi, M., \& Chieffi, A.\ 2012, \apjs, 199, 38 


\bibitem[Lodders(2003)]{Lodders} Lodders, K.\ 2003, \apj, 591, 
1220 

\bibitem[Moriya et al. (2010)]{moriya10} Moriya T., Tominaga N., Tanaka M., Nomoto K., Sauer D.N., Mazzali P.A., Maeda K., Suzuki T., 2010, \apj, 719, 1445

\bibitem[Nissen et 
al.(2007)]{nissen07} Nissen, P.~E., Akerman, C., Asplund, M., et al.\ 2007, \aap, 469, 319 

\bibitem[Pinsonneault(1997)]{pinson} Pinsonneault, M.\ 1997, \araa, 35, 557 

\bibitem[Podobedova et al.(2009)]{podobedova} Podobedova, L.~I., 
Kelleher, D.~E., 
\& Wiese, W.~L.\ 2009, Journal of Physical and Chemical Reference Data, 38, 171 

\bibitem[Salvadori et al.(2007)]{Salvadori} Salvadori, S., 
Schneider, R., \& Ferrara, A.\ 2007, \mnras, 381, 647 

\bibitem[Sbordone et 
al.(2010)]{sbordone} Sbordone, L., Bonifacio, P., Caffau, E., et al.\ 2010, \aap, 522, A26 



\bibitem[Schneider et al.(2003)]{Schneider03} Schneider, R., 
Ferrara, A., Salvaterra, R., Omukai, K., \& Bromm, V.\ 2003, \nat, 422, 869 

\bibitem[Schneider et al.(2012a)]{Schneider12} Schneider, R., 
Omukai, K., Bianchi, S., \& Valiante, R.\ 2012a, \mnras, 419, 1566 

\bibitem[Schneider et al.(2012b)]{Schneider12b} Schneider, R., 
Omukai, K., Limongi, M., et al.\ 2012b, \mnras, L444 

\bibitem[Spite et 
al.(2005)]{spite05} Spite, M., 
Cayrel, R., Plez, B., et al.\ 2005, \aap, 430, 655 


\bibitem[Spite et 
al.(2011)]{spite11} Spite, M., Caffau, E., Andrievsky, S.~M., et al.\ 2011, \aap, 528, A9 


\bibitem[Takeda et al.(2002)]{takeda02} Takeda, Y., 
Parthasarathy, M., Aoki, W., et al.\ 2002, \pasj, 54, 765 




\bibitem[Takeda et al.(2005)]{Takeda} Takeda, Y., Hashimoto, 
O., Taguchi, H., et al.\ 2005, \pasj, 57, 751 



\bibitem[Umeda 
\& Nomoto(2003)]{Umeda} Umeda, H., \& Nomoto, K.\ 2003, \nat, 422, 871 


\bibitem[Van Winckel et al.(1992)]{vanwinckel} Van Winckel, H., 
Mathis, J.~S., \& Waelkens, C.\ 1992, \nat, 356, 500 






\bibitem[Venn 
\& Lambert(2008)]{VennLambert} Venn, K.~A., \& Lambert, D.~L.\ 2008, \apj, 677, 572 


\bibitem[Waelkens et al.(1991)]{waelkens} Waelkens, C., Van Winckel, H., Bogaert, E., \& Trams, N.~R.\ 1991, \aap, 251, 495 




\bibitem[{{Wedemeyer} {et~al.}(2004){Wedemeyer}, {Freytag}, {Steffen},
  {Ludwig}, \& {Holweger}}]{Wedemeyer2004A&A...414.1121W}
{Wedemeyer}, S., {Freytag}, B., {Steffen}, M., {Ludwig}, H.-G., \& {Holweger},
  H. 2004, \aap, 414, 1121


\bibitem[Zhang et al.(2008)]{zhang} Zhang, W., Woosley, 
S.~E., \& Heger, A.\ 2008, \apj, 679, 639 




\end{thebibliography}
\end{document}